\documentclass[aps,amsmath,amyssymb,reprint]{revtex4-1} 
\usepackage{graphicx}

\usepackage{amsmath}
\usepackage{bm}

\usepackage{siunitx}

\usepackage{enumitem} 

\begin{document}
\title{Calculating polaron mobility in halide perovskites}

\date{\today}

\author{Jarvist Moore Frost}
\affiliation{Department of Materials, Imperial College London, Exhibition Road, London SW7 2AZ, UK}
\email{jarvist.frost@imperial.ac.uk}
\affiliation{Centre for Sustainable Chemical Technologies and Department of Chemistry, University of Bath, Claverton Down, Bath BA2 7AY, UK}


\begin{abstract}
%
    Lead halide perovskite semiconductors are soft, polar, materials. 
    The strong driving force for polaron formation (the dielectric electron-phonon
    coupling) is balanced by the light band effective-masses, leading to
    a strongly-interacting large-polaron. 
    A first-principles prediction of mobility would help understand the
    fundamental mobility limits. 
    Theories of mobility need to consider the polaron (rather than
    free-carrier) state due to the strong interactions. 
    In this material we expect that at room temperature polar-optical phonon
    mode scattering will dominate, and so limit mobility. 
    We calculate the temperature-dependent polaron mobility of hybrid halide
    perovskites by variationally solving the Feynman polaron model with the
    finite-temperature free-energies of \=Osaka. 
    This model considers a simplified effective-mass band-structure interacting
    with a continuum dielectric of characteristic response frequency. 
    We parametrise the model fully from electronic-structure calculations. 
    In methylammonium lead iodide 
    at \SI{300}{\kelvin} we predict electron and hole mobilities of
    \num{133} and \SI{94}{\centi\metre\squared\per\volt\per\second}
    respectively. 
    These are in acceptable agreement with single-crystal measurements,
    suggesting that the intrinsic limit of the polaron charge carrier state has
    been reached. 
    Repercussions for hot-electron photo-excited states are discussed.
    As well as mobility, the model also exposes the dynamic structure of the
    polaron.  
    This can be used to interpret impedance measurements of the charge-carrier
    state. 
    We provide the phonon-drag mass-renormalisation, and scattering time constants.
    These could be used as parameters for larger-scale
    device models and band-structure dependent mobility simulations.

\end{abstract}

\maketitle

\section{Introduction\label{introduction}}

Hybrid halide perovskites are a new class of solution processed
semiconductors\cite{Manser2016}. 
They are of considerable practical interest, mainly due to the high
photovoltaic action and potential low cost. 
One important figure of merit for a semiconductor is the charge-carrier
mobility. 

The material is polar and soft. 
This leads to a large dielectric electron-phonon coupling. 
Bare charge-carriers are dressed in phonon excitations, forming a polaron. 

%
In this work we calculate the temperature-dependent polaron mobility. 
We write custom codes to variationally solve the Feynman polaron
model\cite{Feynman1955} for finite temperature\cite{Osaka1959}. 
The inputs to this model are fully specified by electronic-structure
calculations, with no free or empirical parameters. 
We then solve for low-field temperature-dependent mobility with the
FHIP\cite{Feynman1962}, Kadanoff\cite{Kadanoff1963} and Hellwarth et
al.\cite{Hellwarth1999} methods. 

In a photovoltaic device architecture, the mobility limits the thickness of the active
layer. 
In a laser diode, carrier-mobility suggests whether population-inversion (and thus
lasing) is feasible. 
%
Different measures of mobility have different systematic errors, probe
different charge carrier densities, are only sometimes selective to hole and
electron mobilities, and pose general experimental challenges that limit
temperature-dependent (cryostat) measurements. 
To measure intrinsic mobilities, considerable technical effort must be made in
sample preparation, to create sufficient pure and crystalline materials.

Methylammonium lead iodide perovskite (MAPI) is the most well studied of the
halide perovskites. 
Time-resolved microwave-conductivity (TRMC) is a contactless method of
measuring mobility, suitable for temperature dependent measurements. 
Measures on polycrystalline MAPI give a room-temperature mobility-yield product of 
\SI{35}{\centi\metre\squared\per\volt\per\second}\cite{Milot2015}. 
A more recent measure of a single-crystal gives
\SI{115}{\centi\metre\squared\per\volt\per\second}\cite{Semonin2016}. 
It is not known a priori whether these are the maximum achievable, or which scattering process limits the mobility. 


A fully ab-initio method to predict mobilities for infinite perfect crystals
would help understand the implicit limitations of a new material.  

From an electronic structure calculation, one can fit the dispersion relation
(energy vs crystal momentum) for the band extrema to a quadratic function. 
The curvature of this quadratic is the bare-electron band effective-mass. 
This describes a quasi-particle for an singly-charged excitation in the
interacting electron system, with a mass renormalised relative to that of the free electron. 
In lead halide perovskites proper inclusion of spin-orbit-coupling leads to
a complex and non-quadratic dispersion relation\cite{Brivio2014}. 
Nevertheless effective-masses of of 0.12 $m_e$ and 0.15 $m_e$ can be
ascribed\cite{Brivio2014} to electrons and holes respectively ($m_e$ is the
bare-electron mass). 

Continuous thermal disorder interacting with spin-orbit coupling further
complicates this picture.  The band-structure is a dynamic and spin-dependent
object, responding to local fluctuations in electric field\cite{Azarhoosh2016}. 
The lattice dynamically responds to the presence of a charge-carrier,
perturbing the electronic structure of the orbitals in which the charge carrier
sits. 
The full treatment of charge-carrier mobility in these materials is clearly
complex, and well beyond what has been attempted so far. 
Given the dynamic nature of the electronic structure, it is not at all clear
whether there is anything to be gained by including the rigid-band structure of
a particular stochastic realisation of a disordered material, in a mobility
calculation.

The bare-electron rigid-band effective-mass is an inertia term. 
Irrespective of the effective-mass, in the absence of scattering, mobilities
are infinite. 
Yet a small effective-mass (highly disperse bands) is often used as
a predictor of a high mobility. 
This correlation assumes that the scattering processes are independent of (or
at least weakly correlated with) the effective-mass.  
This is the constant scattering time approximation. 
Small effective-masses therefore correlate with the charge-carrier accelerating
to a higher velocity in the time before relaxation events. 
This scattering time has to be inserted into the theory, and is often used as a free parameter. 
For hybrid halide perovskites, in analogy with the similar dispersion relation
and effective mass present in CdTe, and assuming typical covalent semiconductor
scattering times, a mobility as high as
\SI{1000}{\centi\metre\squared\per\volt\per\second} would be inferred. 
This is a considerable overestimate compared to what has been observed.

As the material is soft, the Debye temperature for low-energy optical modes in
this system is $\approx$\,\SI{110}{\kelvin} (\SI{2.25}{\tera\hertz}
\cite{Brivio2015}). 
Charge carriers above this temperature can scatter inelastically by emitting
optical phonons\cite{Ridley5thEd}, a highly dissipative process. 
Impurity and acoustic-phonon scattering are elastic, so less dissipative. 
They would be expected to dominate mobility below this threshold temperature.

The same processes are at play in a standard covalent
semiconductor\cite{Ridley5thEd}, but the relevant temperatures are shifted in
in proportion to the stiffening of the phonon modes. 
For example, in GaP the mobility above \SI{400}{\kelvin} is limited by
optical-mode scattering\cite{Rode1972}. 
The Debye temperature is \SI{580}{\kelvin} for the \SI{12.06}{\tera\Hz} LO mode
of GaP\cite{Pdr1983}).  

There is evidence that anti-bonding frontier orbitals (equivalently, 'inverted'
band structure) in the material class\cite{Huang2013} leads to
a 'defect-tolerance', with defect states repelled away from the
gap\cite{Brandt2015}. 
With fewer electrically-active impurities there is less impurity scattering. 
The large static dielectric constant also shields charged impurities, reducing the cross-section for scattering. 

Though crystalline, halide perovskites are polar and soft. 
This leads to a large zero-frequency (static) dielectric constant. 
The lattice deforms around a charge carrier, localising it. 
This provides a large driving force for stabilising the polaron. 

The halide perovskites are unusual in that they are highly polar yet possess
light effective masses.  
There is a large dielectric electron-phonon interaction, yet the kinetic energy
of the electron is sufficient to keep the polaron from collapsing into
a localised (small-polaron) state. 
A correct transport theory in this material must incorporate the strongly
interacting nature of the polaronic charge-carriers.  
Previous theoretical studies of mobility in this
material\cite{Motta2015,Zhao2016,Filippetti2016} have generally solved the Boltzmann
equation in the relaxation time approximation, with no explicit treatment of
the polaron state. 


In this work we return to direct theories of polaron mobility. 
We develop general codes to apply these methods to arbitrary polar systems, and
present results for halide perovskite materials. 
We predict the temperature dependence and absolute value of
polaron mobility in halide perovskites, without empirical parameters.
The system we consider is highly idealised. 
The electronic band-structure is present only as an effective mass.  
The physical response of the lattice is parametrised by optical and static
dielectric constants, and an effective dielectric-response phonon frequency.
We only consider the polaron state, and its scattering with this characteristic phonon. 
As such, the calculated mobilities are an upper bound for a perfect single crystal. 

We predict the same temperature dependence of mobility as that shown by
time-resolved microwave conductivity of a polycrystalline
sample\cite{Milot2015}. 
Good absolute agreement is found with (room-temperature only) single crystal
terahertz conductivity measurements\cite{Saidaminov2015,Semonin2016}.

In our model we ignore the perovskite phase transitions present in the system,
assuming a constant effective mass and phonon response. 
We do not consider additional scattering from dissipative rearrangement of the
ions, which may be particularly relevant for the (dynamic) cubic phase. 

Our calculations explain the published data---the model has predictive
power. 
This suggests that the intrinsic performance limits of the material have been
realised, and that polaron optical-phonon scattering dominates room temperature
mobility. 

As well as predicting the mobility (a phenomenological quantity), we
characterise the nature of the polaron state, calculating phonon-drag mass
renormalisation and scattering rates, and a polaron size.
These can complement experiments to characterise the charge carrier state. 
In particular, Hendry et al.\cite{Hendry2004} applied the same models we use
here to polaron mobility, effective mass, and scattering time in TiO$_2$.

Further measurements of temperature-dependent mobility in the halide
perovskites will help understand the charge-carrier state and scattering
processes in these materials. 

\section{Methods\label{methods}}

\subsection{Electron phonon coupling}

The interaction between charge-carriers and lattice vibrations is mediated by
the electron-phonon coupling. 
This is challenging to calculate by first principle\cite{Giustino2017}. 
A common method is to use density-functional perturbation-theory. 
This should continue the greater contribution for a hard (covalent) system,
where the atomic motion is small. 
For a soft system such as hybrid perovskites, this lattice distortion
contribution is large.
Non-perturbative methods may be necessary\cite{Whalley2016,Saidi2016}.
Within perturbation theory, there is uncertainty about which diagrams (types
and orders of perturbative interaction) to include in the summation. 
The spiky nature of the numerical integration across the double (phonon and
electron) Brillouin zones makes the calculations heavy and convergence
difficult.
 
For polar systems, the major contribution to electron-phonon coupling comes
from the long-ranged electric fields generated by the atomic displacements. 
This is specified by the dielectric response of the material.  
This considerably simplifies the problem. 

Fr\"ohlich\cite{Frohlich1952} first constructed a Hamiltonian for a system of
independent (i.e. low density) electrons interacting with harmonic
(non-interacting) polar optical phonons. 
The dimensionless Fr\"ohlich parameter $\alpha$ of dielectric electron-phonon
coupling is 

\begin{equation}
\alpha = 
    \frac{1}{4\pi\epsilon_0} 
    \frac{1}{2} 
\left ( \frac{1}{\epsilon_{\infty}}-\frac{1}{\epsilon_S} \right )
\frac{e^2}{\hbar\Omega}
\left( \frac{2m_b \Omega}{\hbar} \right)^{1/2}
.
\end{equation}

This is fully defined by the material specific properties of the optical
($\epsilon_{\infty}$) and static ($\epsilon_S$) dielectric constants, the
bare-electron band effective-mass ($m_b$), and a characteristic phonon angular
frequency ($\Omega$).  
As usual, $2\pi\hbar$ is Planck's constant, $\epsilon_0$ the permittivity of
free space, and $e$ the electron charge. 

The optical and static dielectric constants form a pre-factor for this
electron-phonon interaction. 
This is the response of the lattice, modelled as a continuum dielectric,
beyond the boundary of the polaron. 
Within the polaron, only the electronic excitations (optical dielectric
constant) can keep up with the motion of the electron. 
Outside the boundary of the polaron, the polar lattice excitations can also
respond.

\subsection{Multiple phonon branches}

The static dielectric constant can be calculated by a summation
over the Brillouin zone centre (gamma point) harmonic phonon modes. 
Only these modes contribute, as they have no phase factor which otherwise leads
to a zero contribution when integrating over real-space. 
These phonon mode eigenvectors are used to project the Born
effective charges, to give an effective dipole. 
This is also the infrared activity of the mode. 
Integrating through these individual Lorentzian responses leads to the
dielectric function\cite{Gonze1997}, 

\begin{equation}
\epsilon_S = 
\epsilon_{\infty} +  
    \frac{4\pi}{V}
    \sum_{i=0}^{N} \frac{(Z^*U^*_i)(Z^*U_i)}{\Omega_i^2}
.
\end{equation}

Here, the summation is over the $N$ phonon modes, normalised by $V$ the
unit-cell volume. 
$Z^*$ are the Born effective charges, $U_i$ a specific phonon eigenmode,
$\Omega_i$ the phonon frequency. 

The factor of $\Omega_i^{-2}$ means the summation is dominated by the lowest
energy (frequency) modes which are infrared-active (polar). 
In the case of a two-atom unit cell such as most covalent semiconductors, there
is one infrared-active mode, the linear-optical (LO) mode. 
Most of the polaron literature is constructed in this framework of a single
polar response frequency. 

Hellwarth et al.\cite{Hellwarth1999} provide a prescription to reduce multiple
infrared-active phonon branches to a single equivalent dielectric-response
phonon frequency. 

%

\subsection{Feynman polaron model}

Feynman solved the Fr\"olich Hamiltonian in an innovative manner\cite{Feynman1955},
where the electron interacting with a cloud of independent (harmonic) phonon
excitations is path-integrated over the (phonon) quantum field. 
The electron interacts with the disturbance it has previous created in passing
through the lattice, which exponentially dies out in time. 
This, Coulomb-like, interaction would be expected to depend on inverse-distance. 
Instead, a harmonic model is constructed. 
This can be analytically path-integrated. 
An exponential dampening factor $w$ adds a degree of freedom for
anharmonicity, and the interaction path-integral is scaled by a coupling
strength $C$. 

These model parameters are then varied to minimise the (athermal)
ground state energy, for a given coupling strength $\alpha$ and phonon
response frequency $\Omega$. 

The resulting model is a single particle system where the electron interacts
with a time-retarded potential. 
The Hamiltonian in the centre-of-mass frame can be expressed as an electron
coupled to a finite mass ($M$, expressed in units of the charge carrier
effective mass) with a (harmonic) spring-constant ($k$).  
These give rise to an angular-frequency for the oscillation of the energy
between the electron and phonons of $w=\sqrt{\frac{k}{M}}$.  

It is convenient to work in this angular-frequency variable and make
a further substitution of $v^2=w^2+\frac{4C}{w}$. 
Here $C$ is the electron-phonon coupling coefficient of the model Lagrangian,
related back to the spring-constant by $k=\frac{4C}{w}$\cite{Feynman1955}. 

Feynman's model is non-perturbative and so correct to all orders in the coupling
constant $\alpha$. 
Up until this point, theories had either been based on an assumption of weak
coupling ($\alpha<<1$) and perturbative, or assumed strong coupling
($\alpha>10$). 
Many systems of experimental interest have an intermediate coupling. 
We will show later that inorganic halide perovskites have a characteristic
$\alpha$ coupling of 1 to 2, whereas the additional dielectric response of the
molecular cation gives the hybrid halide perovskites an $\alpha$ of 2 to 3.

\=Osaka\cite{Osaka1959} extended Feynman's athermal variational solution by
providing finite-temperature free-energies of the coupled electron-phonon
system. 
The parameters can then be varied to minimise the total polaron free-energy (at
specific temperature).  
The model thus becomes explicitly temperature dependent, rather than using the
athermal variational solution with a temperature-dependent mobility theory. 
Here we use a more recent presentation of the \=Osaka free-energies by Hellwarth
et al.\cite{Hellwarth1999}, which have been made more amenable to numeric
computation. 

\subsection{Polaron mobility}

By itself, the variational (finite-temperature extended) Feynman model provides
an effective-mass of the polaron quasi-particle representing the phonon-drag
term, and a spring coupling constant for the exchange of energy between the
electron and its coupled phonon cloud. 
As with a bare-electron band effective-mass, there is no dissipative
(frictional) term. 
To calculate a polaron mobility, we must analyse the response of this system to
a perturbation. 
As our main concern is with application to photovoltaic materials, we are
interested in the small-field (direct current, DC) limit. 

The models used herein consider the electronic band structure only in an
effective-mass approximation. 

The initial mobility work of FHIP\cite{Feynman1962} directly used the
response of the polaron centre-of-mass coordinate to an applied field, containing
interactions with a dynamically maintained steady state of thermally excited
phonons. 
A general expression for the impedance is given therein (equations 46 and 47). 
A low temperature approximation is made (a power-series expansion in the
internal parameter $b$, small for low-temperature), leading to a mobility
generally known as $\mu_{FHIP}$, 

\begin{equation}
    \mu_{FHIP} = 
        \left(\frac{w}{v}\right)^3
        \frac{3 e}{4 m_b}
        \frac{exp\left(\beta\right)}{\Omega\alpha\beta} 
        exp\left(\frac{v^2-w^2}{w^2v}\right)
        .
\end{equation}

Here $w$ and $v$ are the (variational) parameters specifying the polaron model,
$m_b$ is the bare-electron band effective-mass, $\Omega$ the phonon angular
frequency, and $\beta=\hbar\Omega/k_BT$ is a reduced thermodynamic temperature
in units of the phonon energy. 

This formalism was quickly realised to be pathological for high
temperatures---at the thermal energy ($k_b T$) equaling the phonon energy
($\hbar\Omega$) there is a resonance minimising the mobility, but the mobility
thereafter increases as a function of temperature. 
Soon afterwards, Kadanoff\cite{Kadanoff1963} provided a similar mobility
identity based on a solution to the Boltzmann equation, applicable to all
$\alpha$. 
This Boltzmann solution implicitly assumes independent scattering events,
still formally limiting the model to low temperature. 
We refer to this as $\mu_K$,

\begin{equation}
    \mu_{K} = 
        \left(\frac{w}{v}\right)^3 
        \frac{e}{2 m_b}
        \frac{exp\left(\beta\right)}{\Omega\alpha} 
        exp\left(\frac{v^2-w^2}{w^2v}\right)
        .
\end{equation}

Though derived from different assumptions, compared to the FHIP
there is a additional factor of $3/2\beta$. 
This is now understood\cite{Peeters1983} to formally arise from the order in which
the limits are taken in the asymptotic approximations. 
Physically, the FHIP considers stimulated-emission and absorption of phonons,
which saturates and balances, once there is a thermal population (Bose-Einstein
statistics). 
The Kadanoff identity adds spontaneous emission of phonons. 
Empirically this corrects the pathological behaviour of FHIP for temperatures
above the Debye temperature, the Kadanoff mobility asymptotically
approaching a constant. 

The relaxation-time approximation (independent scattering) used by Kadanoff
allows a direct evaluation of a scattering rate, based on the thermal
population of phonons $\bar N = exp(\beta)$. 
The rate ($\Gamma$) is evaluated for small momentum exchanges (low-fields) as 

\begin{equation}
    \Gamma_0 = 2 \alpha\bar N \sqrt{(M+1)} exp(-M/v) .
\end{equation}

Here $M$ is the phonon-drag effective-mass (in units of the band
effective-mass, $m_b$). 

This rate $\Gamma_0$ is expressed in reduced phonon units. 
Multiplying with the phonon frequency $\frac{\Omega}{2\pi}\Gamma_0$ gives
a real-time rate. 
This scattering rate can be directly related to $\mu_{K}$ (as it is assumed to
follow a Boltzmann equation) by 

\begin{equation}
    \mu_K = \frac{e}{\Omega m_b (M+1) \Gamma_0 }. 
\end{equation}

More recently, Hellwarth et al.\cite{Biaggio1997,Hellwarth1999} returned to the
general response theory of FHIP (Equations 46 and 47 in \cite{Feynman1962}). 
Rather than taking a low-temperature limit and perform a power-expansion, they
contour integrate for the self-energy of the perturbed polaron. 

\begin{equation}
    a^2 = 
        (\beta/2)^2 + 
        \left(\frac{v^2-w^2}{w^2v}\right) 
        \beta coth(\beta v/2)
    , 
\end{equation}

\begin{equation}
    b = 
        \left(\frac{v^2-w^2}{w^2v}\right) 
        \frac{\beta}{sinh(\beta v/2)}
    , 
\end{equation}

\begin{equation}
    K = 
    \int_{0}^{\infty} du 
        \left( u^2 + a^2 - b cos(vu) \right) ^{-\frac{3}{2}} cos(u)
    .
\end{equation}


This integral gives the polaron response to a first order change in the driving
force, providing an analogous role to the scattering rate in Kadanoff's
Boltzmann construction, to give, 
\begin{equation}
    \mu_{H}  = 
    \left(\frac{w}{v}\right)^3 
    \frac{3 e}{m_b}
    \frac{\sqrt{\pi} sinh(\beta/2) }{ \Omega \alpha\beta^{\frac{5}{2}} }
    K^{-1}
   .
\end{equation}

This is as in reference \cite{Hellwarth1999}, transformed into S.I. units and
rearranged for easier comparison with $\mu_{FHIP}$ and $\mu_K$.

Both Biaggio et al.\cite{Biaggio1997} and Hellwarth et al.\cite{Hellwarth1999}, approximate
$b=0$, allowing for an analytic solution of $K$ with modified Bessel functions. 
Here we do the full integration. 
We note that both papers have a typographic error in the formula
for $b$, possessing a spurious term of $b$ on the right hand side. 
The form here is as given by FHIP\cite{Feynman1962}. 

In custom codes we reimplement the Hellwarth et al.\cite{Hellwarth1999} posing of
\=Osaka's\cite{Osaka1959} finite-temperature variational solution to
Feynman's\cite{Feynman1955} model. 
Integration is numeric using an adaptive Gauss-Kronrod quadrature algorithm. 
The \=Osaka free-energies (which include numeric integrals) are automatically
differentiated in a forward-mode to produce gradients. 
These gradients are used by a BFGS algorithm to find the optimal
(temperature-dependent) $v$ and $w$. 
These model parameters directly give the FHIP\cite{Feynman1962} and
Kadanoff\cite{Kadanoff1963} mobility. 
By performing the contour integration for the polaron self-energy numerically,
we calculate an Hellwarth et al.\cite{Hellwarth1999} mobility (with
a slight refinement by considering $b\neq0$).
Codes are provided\cite{GitHub} to encourage the application of these methods
to other systems of interest. 

\section{Results\label{results}}

\subsection{Methylammonium lead iodide perovskite}

Methlyammonium lead halide is the most well studied of the hybrid halide
perovskites. 
We use the previously mentioned QS\textit{GW} effective masses of 0.12
(electron) and 0.15 (hole)\cite{Brivio2014}. 
We take $4.5$ and $24.1$\cite{Frost2014} as the optical (QS\textit{GW}) 
and static (harmonic phonon, DFT) dielectric constants. 

We note that this static dielectric constant only includes the harmonic
response of the phonons, additional (slower) terms may come from the
(anharmonic) realignment of the polar methylammonium. 
We can estimate this contribution from a generalised form of Onsager theory for
polar liquids\cite{Kirkwood1939}. 
Ignoring the response of the local environment and assuming the ions totally
free to realign (i.e. most applicable to the cubic phase), this is $\epsilon_p
= \frac{4}{3}\pi N \frac{\mu^2}{3kT}$.  
With the methlyammonium-dipole $\mu=2.2$D \cite{Frost2014}, and $N$ the number
density (one methylammonium per \SI{6.2}{\angstrom} cubic unit cell), this
gives $\epsilon_p = +8.9$.

It is well known that the increasing frustration of motion of these dipoles as
the material passes through the second-order tetragonal phase transition
towards the fully hindered orthorhombic ground state gives rise to a divergent
dielectric response\cite{OnodaYamamuro1992}. 
It is not clear how to integrate this (slow, dissipative) response 
into a formal polaron transport theory. 

One might assume that the additional response in the cubic phase, and in the
divergence approaching the orthorhombic phase, would provide an additional
dissipation of electron energy, and so lower mobility.  
We ignore these terms in the present work. 
Disorder in the hybrid material will generate further
localisation pressure, perhaps even forming small
polarons\cite{Ma_2015,Neukirch_2016}. 

Hellwarth et al.\cite{Hellwarth1999} provides two approximation schemes for reducing multiple
phonon branches in the polaron problem to an single dielectric-response frequency. 
In this work we use the more simple athermal 'B' scheme. 
The temperature-dependent 'A' scheme may offer more temperature-dependent
phonon structure in the polaron model, and therefore the resulting
temperature-dependent mobilities. 
This will be the subject of future work. 

We take our modes from ab-initio phonon and infrared-activity
calculations\cite{Brivio2015}. 
Applying the 'B' scheme to all modes (including the 18 high frequency molecular
modes) gives a seemingly unphysical characteristic response of \SI{10.0}{\tera\hertz}. 
It is not clear whether this is due to errors in our estimates of the infrared
activity of the molecular modes, or whether the Hellwarth et al.
summation is only well defined for long range consummate lattice distortions. 

Applied to the first 15 (non-intramolecular) modes, the effective
dielectric-response ('LO') phonon frequency is \SI{2.25}{\tera\hertz}. We use
this value in all work presented herein. 

Combined, these data specify the Fr\"ohlich electron-phonon parameter as
$\alpha=2.40$ (electron) and $\alpha=2.68$ (hole). 

Bokdam et al.\cite{Bokdam2016} provide a direct evaluation of the total
dielectric function. 
From a density functional perturbation theory calculation (harmonic athermal
response), the imaginary dielectric function has a peak at \SI{8}{\meV}
(\SI{1.9}{\tera\hertz}). 
This corresponds well with the estimate from the Hellwarth et al. B scheme
applied to the lattice modes.
Including some anharmonic contributions via ab-initio molecular dynamics, they
find a softening of the modes to \SI{4}{\meV} at \SI{300}{\kelvin}. 
This implies that polaron effects are actively strengthened at higher
temperature, as the anharmonic softening of the lattice increases the effective
dielectric electron-phonon coupling. 

\begin{figure}
    \includegraphics[width=1.0\columnwidth]{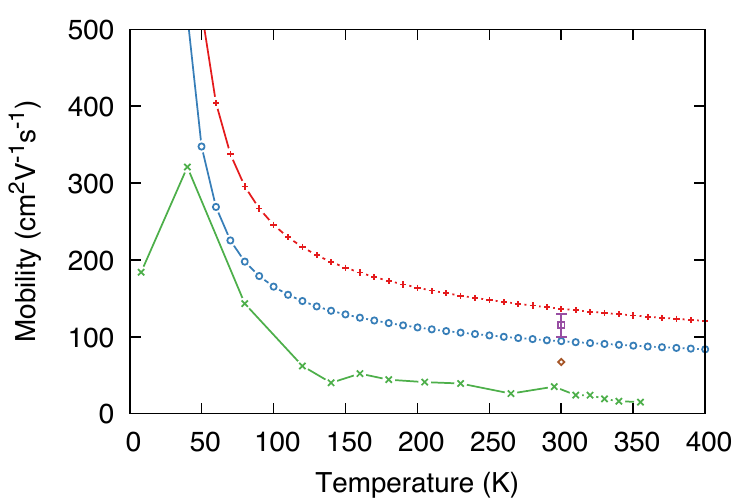}
    \caption{\label{fig-mobility-calculated-experimental}
    Predicted Hellwarth hole (cross, red) and electron (circle, blue) polaron mobility versus temperature,
    co-plotted against temperature-dependent
    time-resolved-microwave-conductivity measurements on polycrystalline films
    from Milot et al.\cite{Milot2015} (green, saltire), room temperature
    space-charge-limited-current hole-mobility estimate on a single crystal
    from Saidaminov et al.\cite{Saidaminov2015} (diamond, brown), and room temperature
    time-resolved-microwave-conductivity on a single crystal Semonin et
    al.\cite{Semonin2016} (square, purple). 
    As the TRMC data is a mobility-yield product, the low \SI{8}{\kelvin} data
    point may be related to the stability of the exciton (none-unitary free
    carrier yield) at those temperatures, rather than the mobility actually
    reducing. 
    }
\end{figure}

At 300 K we predict a Kadanoff polaron mobility of
$\mu_e =$\SI{197}{\centi\metre\squared\per\volt\per\second}, 
$\mu_h =$\SI{136}{\centi\metre\squared\per\volt\per\second};
and a Hellwarth mobility of
$\mu_e =$\SI{133}{\centi\metre\squared\per\volt\per\second}, 
$\mu_h =$\SI{94}{\centi\metre\squared\per\volt\per\second}. 
The phonon mass renormalisation is $0.37$ for the electron polaron and $0.43$
for the hole polaron.
This agrees well with the athermal perturbative (small
$\alpha$) estimate $m_p^*=\frac{\alpha}{6} + \frac{\alpha^2}{40}$
\cite{Feynman1955} which gives 0.54 and 0.63 respectively. 

\begin{table}
\caption{\label{tab:Params}Parameters of the Feynman polaron model as used
    in this work.  
    Relative high frequency ($\epsilon_\infty$) and static ($\epsilon_S$)
    dielectric constants are given in units of the permittivity of free space
    ($\epsilon_0$). Frequency (f) is in \si{\tera\hertz}. Effective mass
    ($m^*$) is in units of the bare electron mass. 
    }
\begin{ruledtabular}
\begin{tabular}{lrrrr}
    Material & $\epsilon_\infty$ & $\epsilon_S$ & $f$ & $m^*$ \\
    \colrule
    MAPbI3-e & 4.5 & 24.1 & 2.25 & 0.12 \\
    MAPbI3-h &     &      &      & 0.15 \\
    CsPbI3   & 6.1 & 18.1 & 2.57 & 0.12 \\
    \colrule 
    MAPbI3  \footnote{These parameters specify the model as in Sendner et
    al.\cite{Sendner2016}. Effective mode frequency, reduced by a Hellwarth et al.
    scheme, is from prviate communication.} 
       & 5.0 & 33.5 & 3.38 & 0.104 \\
    MAPbBr3 & 4.7 & 32.3 & 4.47 & 0.117 \\
    MAPbCl3 & 4.0 & 29.8 & 6.42 & 0.2 \\
    \colrule
    CsSnI3  \footnote{Parameters for holes in cesium tin halides are
    taken from Huang et al.\cite{Huang2013}. Dielectric constants are from Table VII
    therein, effective masses from Table VI, phonon frequency from table VII.}
        & 6.05 & 48.2 &  4.56 & 0.069 \\
    CsSnBr3 & 5.35 & 32.4 &  5.48 & 0.082 \\
    CsSnCl3 & 4.80 & 29.4 &  7.28 & 0.140 \\
\end{tabular}
\end{ruledtabular}
\end{table}

\begin{table}
\caption{\label{tab:Results}Dielectric electron-phonon coupling ($\alpha$,
    athermal). Predicted (\SI{300}{\kelvin}) mobilities (Kadanoff,
    $\mu_{K}$; Hellwarth, $\mu_{H}$), polaron effective mass renormalisation
    ($m_h^*$), Feynman-model variational parameters ($v$ and $w$), Kadanoff
    polaron relaxation time ($\tau$, \si{\pico\second}), and Schultz polaron radius
    ($r_f$, \si{\angstrom}).
    }
\begin{ruledtabular}
\begin{tabular}{llrrccccc}
    Material & $\alpha$ & $\mu_{K}$ & $\mu_{H}$ & $m_h^*$ & $v$ & $w$ & $\tau$ & $r_f$\\
    \colrule
    MAPbI3-e & 2.39 & 197 & 136 & +0.37 & 19.9 & 17.0 & 0.12 & 26.8\\
    MAPbI3-h & 2.68 & 133 &  94 & +0.43 & 20.1 & 16.8 & 0.10 & 25.3\\
    CsPbI3 & 1.35 & 389 & 258 & +0.21 & 16.5 & 15.1 & 0.20   & 39.6\\
    \colrule
    MAPbI3  & 1.71 & 272 & 195 & +0.31 & 13.2 & 11.5 & 0.13 & 43.1\\
    MAPbBr3 & 1.69 & 212 & 157 & +0.36 & 10.1 &  8.7 & 0.12 & 53.6\\
    MAPbCl3 & 2.19 &  73 &  57 & +0.62 &  7.6 &  6.0 & 0.08 & 61.4\\
    \colrule
    CsSnI3  &  1.02 & 703 & 487 & +0.20 & 9.64 & 8.81 & 0.21 & 70.0\\
    CsSnBr3 &  1.09 & 511 & 356 & +0.24 & 8.14 & 7.32 & 0.19 & 77.7\\
    CsSnCl3 &  1.39 & 212 & 147 & +0.36 & 6.51 & 5.59 & 0.14 & 85.4\\
\end{tabular}
\end{ruledtabular}
\end{table}

\begin{figure}
    \includegraphics[width=1.0\columnwidth]{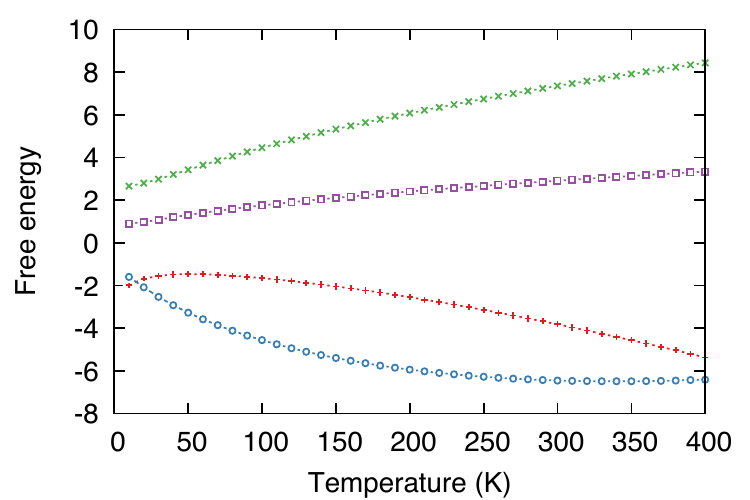}
    \caption{\label{fig-variational}
    Finite-temperature free-energies used in the variational fitting of the $v$
    and $w$ parameters for the coupled electron-phonon system. 
    The upper bound of the free energy (cross, red) is $F=-(A+B+C)$. 
    Here A (circle, blue) is total thermodynamic energy of the system, via the log of the
    partition function of the density matrix. 
    B (saltire, green) and C (squares, purple) are the expectation values of the action
    for the electron-phonon system and the trial action respectively. 
    See reference \cite{Osaka1959}. 
    Data for the electron (0.12) effective mass in MAPI. 
    } 
\end{figure}

\begin{figure}
    \includegraphics[width=1.0\columnwidth]{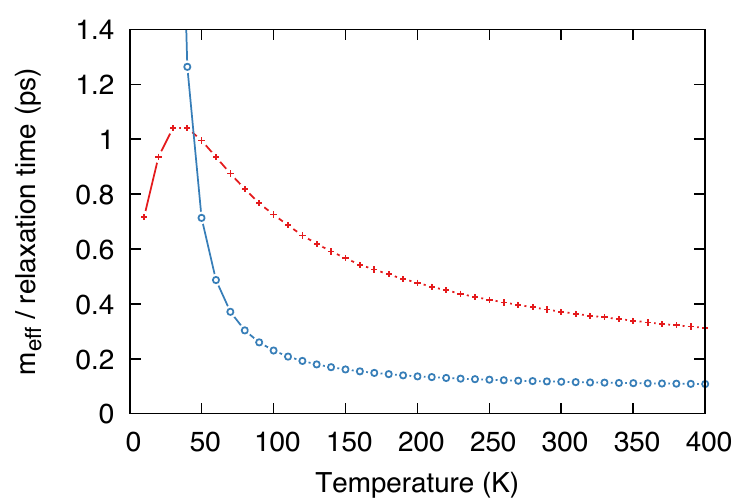}
    \includegraphics[width=1.0\columnwidth]{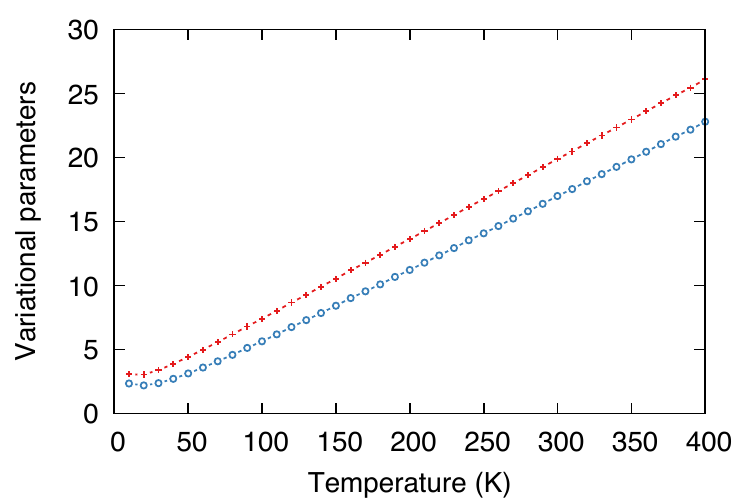}
    \caption{\label{fig-mass-tau}
    (Top) Temperature-dependent phonon-drag effective mass (cross,red) in units of the
    bare-electron band effective-mass. Polaron scattering-time
    (circle, blue) in units of \si{\pico\second}. 
    (Bottom) Temperature-dependent variational parameters $v$ (cross, red) and
    $w$ (circle, blue) for the coupled electron phonon system.
    Data for the electron (0.12) effective mass in MAPI. 
    }
\end{figure}

The variational parameters $v$ and $w$ specify the polaron state. 
These come from minimising the finite-temperature \=Osaka\cite{Osaka1959} free
energies (Figure \ref{fig-variational}). 
These parameters can be mapped to a phonon-drag mass term $M$, and spring
constant $k$, for the effective single-particle coupled harmonic system (Figure
\ref{fig-mass-tau}).

Following Schultz\cite{Schultz1959}, our variational parameters at
\SI{300}{\kelvin} give a polaron radius (for holes and electrons) of
$\approx$\,\SI{26}{\angstrom}. 
(See Table \ref{tab:Results} for further values.)
This radius is defined as the standard deviation of the Gaussian wavefunction which
would form in the harmonic confining potential of the polaron.  
Though the definition is fairly arbitrary, it is of interest in understanding
how commensurate the polaron is to real-space fluctuations in electrostatic
potential.  

As with Hellwarth et al.\cite{Hellwarth1999}, we find that setting b=0 in
our calculation of $\mu_H$ makes only $<0.2\%$ difference to the mobility. 
This difference increases with temperature.

As a function of temperature (Figure
\ref{fig-mobility-calculated-experimental}), the mobility decays from an
infinite quantity at zero temperature. 
The Kadanoff mobility asymptotically approaches 
\SI{190}{\centi\metre\squared\per\volt\per\second} at high temperature.  
This is associated with the variational fit approaching a phonon effective mass
of $0.37 m_e$, while the spring constant $k$ increases linearly with
temperature (Figure \ref{fig-mass-tau}). 
By comparison, the (low-temperature) FHIP mobility has a minimum at
$\hbar\Omega=k_B T$, and increases linearly with temperature. 
The Hellwarth mobility has weak positive temperature dependence at high
temperatures. 

The high temperature behaviour of polaron mobility is important for
photovoltaic device operation. 
Hybrid halide perovskites are predicted to have extremely low
thermal conductivity\cite{Whalley2016}, and cooling of photoexcited states is
known to be unusually slow\cite{Price2015,Yang2015}. 
As such, it is likely that the initial electron temperature of a photo-excited
state is extremely high. 
Optical phonons emitted by electron thermalisation and carrier scattering may
remain in the region of the polaron, reheating the electron state. 
It is therefore useful and interesting to know that even with a high
temperature 'hot carrier', the polaron mobility is finite, and sufficient for
photovoltaic device operation. 

The Kadanoff and FHIP mobilities posess at their core a calculation of
the rate of emission and absorption of optical phonons $\Gamma$, taken to the
low-momentum limit $\Gamma_0$. 
Kadanoff\cite{Kadanoff1963} directly relates this rate to a relaxation-time
approximation of the Boltzmann equation, $\tau=1/\Gamma$. 
We thereby access the scattering time for the polaron as a function of
temperature, finding for the electron-polaron $\tau=$\SI{0.12}{\ps} at
a temperature of \SI{300}{\kelvin}. 
Combined with the polaron mass renormalisation (the phonon drag), these values
parametrise a temperature-dependent relaxation-time approximation Boltzmann
equation for the polarons in this material, and may be of use in larger-scale
device models. 
Further, these parameters can be probed experimentally. 
Previous work\cite{Hendry2004} used Terahertz spectroscopy to probe the
scattering in TiO$_2$, observing a counter-intuitive relationship between
scattering rate and mobility, due to the complementary relationship between
effective-mass and scattering-rate $\mu=e/(m_p \Gamma )$.

\subsection{Powerlaw temperature dependence of mobility}

Inference of the dominant electron scattering mechanism in a material often
comes from the circumstantial behaviour of the mobility as a function of
temperature. 
This temperature-dependent mobility (a phenomenological quantity) often shows
a power-law scaling. 
For the halide perovskites, early data suggests a $T^{-\frac{3}{2}}$ 
exponent, consistent with textbook descriptions of acoustic-mode scattering
dominated mobilities\cite{Ridley5thEd}. 
We previous suggested that activation of multiple soft optical phonon modes
could mimic such (large exponent) behaviour\cite{Leguy2016}. 

\begin{figure}
    \includegraphics[width=0.95\columnwidth]{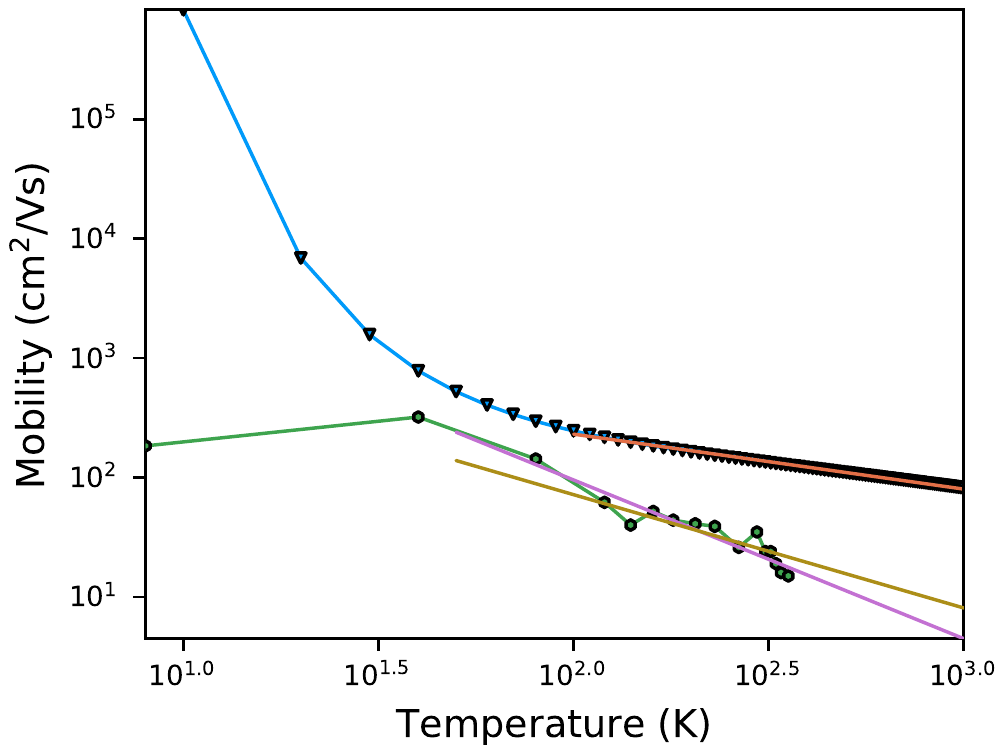}
    \caption{\label{fig-powerlaw}
    Hellwarth electron mobility for MAPI (down-triangles, blue) coplotted on a log-log axis
    with TRMC temperature dependent data of (hexagons, green) of Milot et
    al.\cite{Milot2015}. 
    Fit lines are by a least-squares fit to a power-law ($\mu=aT^k$), in a linear space. 
    Fitting to above \SI{100}{\kelvin}, the predicted Hellwarth mobility (this
    work) has an exponent (orange trend line) of $-0.46\approx-\frac{1}{2}$. 
    Fitting all but the spurious first point of the Milot et al. data produces
    an exponent (purple trend line) of $k=-0.95$, whereas fitting just above
    \SI{100}{\kelvin} finds a lower exponent (brown trend line) of $k=-0.95$. 
    } 
\end{figure}

Do polaron mobility theories follow a power-law scaling, and if so, what is the
critical exponent? 
Displaying the Hellwarth mobility on a log-log scale (Figure \ref{fig-powerlaw}
reveals straight-line (power-law) behaviour, for temperatures above the
critical phonon emission energy. 
Fitting these data to a power-law ($\mu=aT^k$) by least-squares minimisation of
data above \SI{100}{\kelvin} finds an exponent of $k=-0.46$. 
This is close to $-\frac{1}{2}$, which we propose as our best estimate of the
power-law behaviour of polaron optical-phonon scattering dominated mobility,
once above the phonon emission threshold. 
(This will hold even in the case where there are multiple distinct optical
phonon thresholds, as all modes will produce a $k=-\frac{1}{2}$ scaling once
above their emission threshold.)

This exponent is considerly less than the $k=-1.33$ extracted from a full fit
to the TRMC data of Milot et al.\cite{Milot2015}. 
However, if we similarly constrain the experimental fit to above \SI{100}{\kelvin}
(motivated by the uncertainity of measures in the low temperature phase, and to
avoid complications due to multiple phonon scattering routes), we
find an exponent of $k=-0.95$. 
Further temperature-dependent measures of mobility will provide stronger
statistical evidence from which to infer the nature of mobility-limit
scattering in these materials.

\subsection{Comparison to Sendner et al.}

Recently, Sendner et al.\cite{Sendner2016} approached the polaron mobility from
an experimental point of view. 
They fitted four critical points to infrared transmissivity measures of thin-films. 
These provide a simplified phonon spectrum. 
This is then used with a generalisation of the Lyddane-Sachs-Teller relation to
derive dielectric constants.  This assumes that there are well defined linear
and transverse optical modes as would be found in a crystal of cubic symmetry. 
These data were then reduced with a Hellwarth scheme to an effective dielectric-response phonon
frequency (for MAPI) of \SI{3.38}{\tera\hertz} (prviate communication).
Combined with a DFT effective mass (0.104), this was then used with a 
Hellwarth model to derive a room-temperature mobility of 
\SI[separate-uncertainty = true]{200 +- 30}{\cm\squared\per\volt\per\second} 
for MAPI. 
See Table \ref{tab:Params} for a statement of their model, and Table \ref{tab:Results}
for a cross-validation with the codes developed in this work. 
With their parameters and our custom codes, we can predict the
temperature-dependent mobility (Figure \ref{figs/Rob-comparison-COLOUR}). 

\begin{figure}
    \includegraphics[width=1.0\columnwidth]{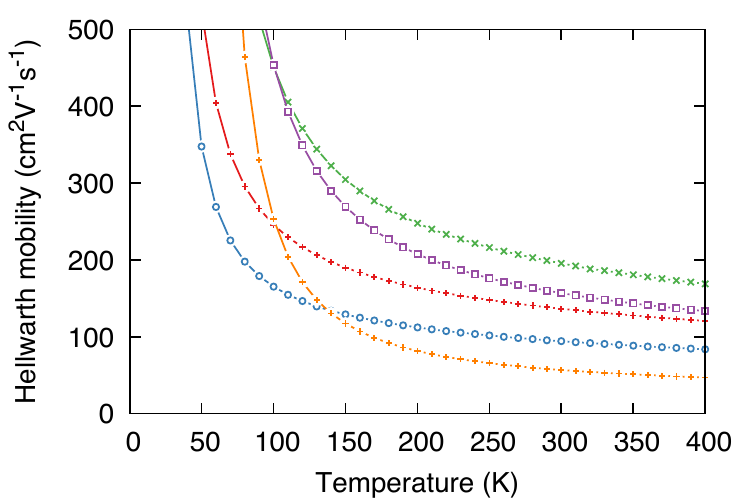}
    \caption{\label{Rob-comparison}
    Calculation of temperature-dependent solutions to the Sendner et
    al.\cite{Sendner2016} (data therein and by private communication)
    parameters, compared to this work. 
    Sendner's parameters for iodine (satire, green), bromine (square, purple)
    and chlorine (cross, orange).
    Data from this work are iodide electrons (cross, red), and holes
    (circle, blue).
    Approximate agreement is found in mobility trends, but due to quite
    different underlying parameters for the polaron state. 
    }
\end{figure}

By comparison to the present work, they extrapolate a greater dielectric
response than we calculate from harmonic phonons. 
This is to be expected as their measurements will include all anharmonicity
(including the molecular rotation contribution), whereas we consider only the
harmonic response. 
However it is not immediately clear that the contribution from this dissipative
molecular response should be integrated into polaron theories built on the
perfectly elastic (non dissipative) response of harmonic phonons.

Though the mobility results of Sendner et al. and this work are in broad
agreement, the underlying parameters of the polaron system disagree. 
Here we suggest a more polaronic state, with a larger dielectric coupling
resulting in a greater electron-mass renormalisation. 
Sendner et al. suggest a less polaronic charge-carrier state, but with
a greater scattering strength due to the larger dielectric constant. 
Overall this leads to a very similar Kadanoff relaxation time (see Table
\ref{tab:Results}).

\subsection{Inorganic lead halide perovskite}

\begin{figure}
    \includegraphics[width=1.0\columnwidth]{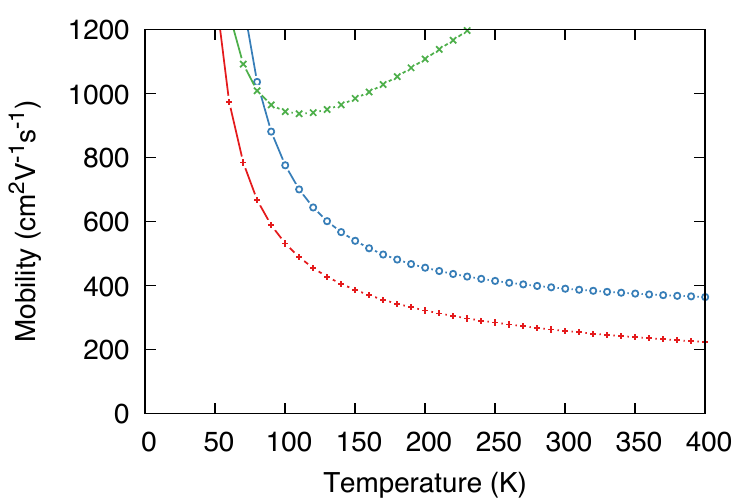}
    \caption{\label{CsPbI3-mobility}
    Predicted mobility for CsPbI3, assuming an effective mass for the carriers
    of 0.12. 
    The theories used are Hellwarth (cross, red), Kadanoff (circle, blue) and
    FHIP (saltire, green). 
    }
\end{figure}

To understand the role of the organic cation in polaron formation and mobility
limits of halide perovskites, we make a comparison with the fully inorganic
analogue, cesium lead iodide. 
We calculate the dielectric constants for cubic cesium lead iodide
(r=\SI{6.78}{\angstrom}), by density functional perturbation theory. 
We use the generalised gradient approximation, in the plane-wave VASP code with
a $9\times9\times9$ Monkhorst-pack k-space integration, \SI{700}{\eV} plane
wave cut-off, and the PBESol functional.  
We find optical and static dielectric constants of $6.1$ and $18.1$. 
The Hellwarth (B-scheme) dielectric effective phonon frequency is slightly
stiffened to \SI{2.57}{\tera\Hz}. 
Assuming the same electron effective mass as electrons in MAPI ($0.12$), this
gives a Kadanoff mobility of 
\SI{389}{\centi\metre\squared\per\volt\per\second}
and a Hellwarth mobility of 
\SI{258}{\centi\metre\squared\per\volt\per\second} at \SI{300}{\kelvin} (see
Figure \ref{CsPbI3-mobility} for temperature-dependence). 

This suggests that the additional (harmonic) dielectric response of the organic ion
in MAPI reduces the mobility by a factor of two, as a result of doubling the
dielectric electron-phonon coupling, leading to a doubling in the polaron
effective mass and a halving of the scattering time. 

\subsection{Inorganic tin halide perovskite}

The cesium tin halide perovskites have been well characterised\cite{Huang2013}
with the QS\textit{GW} method for effective masses, and lattice dynamics with
the LDA density functional.  
Taking values from this publication, we can solve for the polaron model and
mobilities. 
The calculated hole mobilities for CsSnI3 at \SI{300}{\kelvin} are 
\SI{703}{\centi\metre\squared\per\volt\per\second} (Kadanoff) and 
\SI{487}{\centi\metre\squared\per\volt\per\second} (Hellwarth). 
This compares well to measurements of 
\SI{585}{\centi\metre\squared\per\volt\per\second} (Hall-effect) and 
\SI{400}{\centi\metre\squared\per\volt\per\second} (by 'transport property')\cite{Chung2012}. 
Temperature dependence is presented in Figure
\ref{figs/CsSn-halide-mobility-COLOUR}. 

\begin{figure}
    \includegraphics[width=1.0\columnwidth]{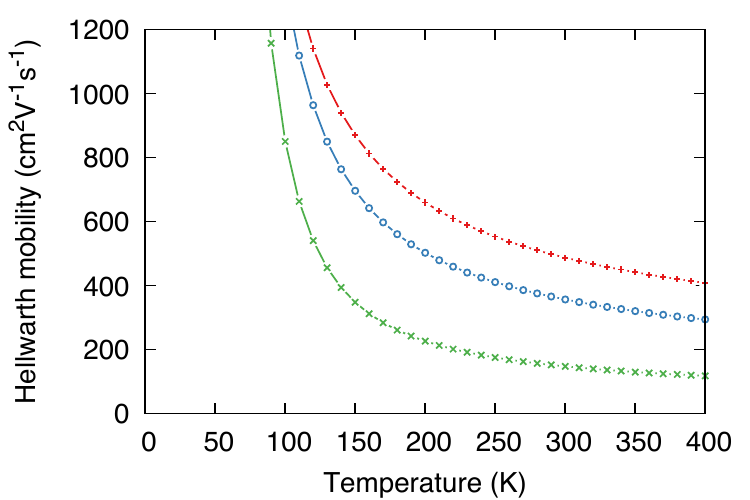}
    \caption{\label{CsSn-mobility}
    Predicted Hellwarth mobility for cesium tin halide perovskites 
    (Iodine: cross, red;
    Bromine: circle, blue;
    Chlorine: saltire, green), 
    assuming an effective mass for the carriers of 0.12.
    }
\end{figure}

In comparison to the lead halide perovskites, the material is less polaronic.
The driving force for polaron stabilisation is greater with the larger
dielectric constants, but the effective masses are smaller, the 
phonon spectrum is stiffer. 
As with MAPI, the agreement between predicted and measured mobility suggests
that polaron mobility describes the charge carrier state and scattering
processes well.

\section{Discussion\label{discussion}}

\subsection{Prior theoretical studies}

Most prior theoretical studies on mobility in hybrid halide
perovskites\cite{Motta2015,Zhao2016,Filippetti2016} have solved the Boltzmann
equation in the relaxation time approximation using the \textsc{BoltzTraP}
codes\cite{Madsen2006}. 

An effective relaxation-time (scattering-constant) $\tau$ is inserted into
a theory otherwise parametrised from electronic structure calculations.  
These works include the band-structure explicitly, whereas we use an effective mass approximation throughout. 
In hybrid halide perovskites, the materials are new and unusual, so there is no
equivalent well characterised system (such as in covalent semiconductors)
on which to base this relaxation time. 

The first work by Motta et al.\cite{Motta2015} used an empirically motivated
scattering time constant of
\SI{1}{\pico\second}. 
An analogy is made by relating this to the time constant of motion of the
methylammonium ion. 
However there is no direct correspondence between the timescale of motion of
the ion and the timescale of charge carrier scattering. 
This led to a predicted mobility of 
\SIrange{5}{12}{\cm\squared\per\volt\per\second} for holes and
\SIrange{2.5}{10}{\cm\squared\per\volt\per\second} for electrons. 
At the time these values were comparable to the highest measured mobilities. 

Later, Zhao et al.\cite{Zhao2016} found a relaxation time of \SI{0.1}{\pico\second} from
acoustic-phonon scattering in a golden-rule formalism with
deformation-potential matrix elements. 
The mobility predicted from this is 
\SI{1000}{\cm\squared\per\volt\per\second}. 
Considering charged-impurity scattering with a static dielectric constant of
6.5, and assuming large non-compensated charge defect densities of
\SI{1e18}{\per\cm\cubed}, these mobilities reduce to 
\SI{100}{\cm\squared\per\volt\per\second}. 
The density of charged impurities and static dielectric constant used in this
work seems unphysical. 

Filippetti et al.\cite{Filippetti2016} combined contributions from impurity
scattering, acoustic-phonon scattering (approximated by the deformation
potential), and a relaxation time approximation recasting of the Fr\"ohlich
optical electron-phonon scattering element. 
They use a static dielectric constant of $60$. 
Overall they predict a 
bare-charge electron scattering time to \SI{9.2}{\femto\second}. 
They find a low charge-density mobility of 
\SI{57}{\cm\squared\per\volt\per\second} (electron) and
\SI{40}{\cm\squared\per\volt\per\second} (hole). 


Very recently Zhang et al.\cite{1708.09469} have provided a semi-classical
model of charge transport in hybrid halide perovskites, based on polarons
scattering from acoustic modes. 
They point out that, by definition, acoustic scattering produces the
experimentally observed $T^{-\frac{3}{2}}$ dependence. 
However they do not consider inelastic scattering processes, such as emission
of optical phonons. 

\subsection{This work}

This work provides the first prediction of (temperature-dependent) polaron
mobility in hybrid halide perovskites. 
There are no free parameters. 

The bare-electron band effective-masses are taken from QS\textit{GW} electronic
structure calculations\cite{Brivio2014}. 
The parameters used in the calculation are the optical and static dielectric
constants, and an effective phonon frequency. 
These values we take from prior density-functional-theory calculations on the
harmonic (phonon) response of the material\cite{Brivio2015}. 

This fully specifies a model Hamiltonian, for which the finite-temperature
Feynman variational solution for the polaron is made. 
Numerical solution of the full DC-mobility theory (without making a Boltzmann
transport equation approximation) provides a temperature-dependent mobility. 


These mobilities are an upper bound, as they consider only one scattering
contribution (that of the polaron scattering with optical-phonons) in an
otherwise perfect effective-mass system. 


Both the Kadanoff and Hellwarth polaron mobility as a function of temperature
show a trend that agrees well with available data.  
As the Kadanoff mobility is based on assuming a Boltzmann equation, this
suggests that the underlying assumption of independent scattering events is
approximately correct.  
The Hellwarth mobility is constructed by an explicit contour integration for the
self-energy of the perturbed polaron.  
We expect it to be more generally applicable, and more accurate for high temperatures. 

The agreement with measured mobilities suggests that we have captured
the essential physics of the system. 

This suggests that impurity scattering is a relatively minor process. 
This adds circumstantial evidence that these semiconductors are 'defect
tolerant', at least as far as mobility is concerned.  
Scattering by acoustic phonon modes is neglected, as it is low-energy and
elastic (due to the thermal population of these low-energy modes). 
Where we predict a divergence to infinite mobility at zero temperature,
acoustic phonon and impurity scattering will come to dominate. 

In these theories we have assumed a single characteristic dielectric phonon frequency.
Though Hellwarth et al.\cite{Hellwarth1999} shows that there are good physical
reasons for the approximation of the action of multiple phonon modes with
a single effective mode, this still loses structure in the
temperature-dependent response. 
In the case of lead-halide perovskites, there are infrared active phonon
branches on \SIrange{1}{1.5}{\tera\Hz} and \SIrange{2}{2.5}{\tera\Hz}. 
Phonon perturbation theory calculations show that phonons in this system are
highly anharmonic\cite{Whalley2016}, with a broad range of energy and small
lifetime. 
This may invalidate the non-interacting (independent) phonon approximation. 

We have shown that the Hellwarth mobility follows an approximately $T^{-0.5}$
power-law dependence, when above the phonon emission threshold. 
More temperature-dependent mobility data will help understand the microscopic
processes responsible for charge carrier scattering in these materials. 

With the robust codes developed for this work, temperature-dependent
large-polaron mobility calculations are made simple. 
The material-specific parameters required from electronic structure
calculations are well defined. 
Modest computational resources are then required to solve the polaron and
mobility theories. 
Unlike other models of halide perovskite mobility presented in the literature,
there is no relaxation-time or other empirical parameter, inferred by analogy
to other systems. 

We have shown that models predict the experimentally observed
temperature-dependent trend in hybrid halide perovskites, and that the
quantitative agreement suggests that polaron optical-phonon scattering
dominates room-temperature mobility. 

\subsection{Future work}

Several classes of novel electronic materials (transparent conducting oxides,
oxide-based thermoelectrics and organic semiconductors) are expected to be
dominated by polaronic transport. 
The calculation of temperature-dependent mobility described in this paper is
useful for both design of new materials, and the interpretation of measurements
and characterisation of existing materials.

Considering the evident utility, it is therefore perhaps surprising that there
has not been more application of these methods. 
Certainly some of this is due to the underlying theoretical work being couched
in the rather esoteric language of path integrals. 
A more practical issue is the lack of an available computational implementation. 
Here we developed custom codes in the Julia high level mathematical
language\cite{Bezanson2017}.
This offered automatic-differentiation (to calculate the gradients of the
objective function of the \=Osaka free-energies, stabilising the optimisation
procedure) and strong control of numeric errors. 
The resulting codes appear numerically stable and general purpose. 
We provide our codes\cite{GitHub} to encourage community reuse of these
methods.

With a validated set of codes, we can now easily apply these polaron mobility
theories to polar materials as a natural component of materials design. 
Though many idealisations are made in the polaron mobility theories, they offer
a first-principles method to arrive at a phenomenological value of extreme
technical interest. 
The electronic structure calculations necessary for inputs to the model are
now a standard part of computational materials discovery and characterisation. 
The methods and codes are well matched to high-throughput calculation of
mobilities. 

Within hybrid perovskites, we plan to combine these models with experimental
measures to characterise the internal structure of the polaron state.
As well as the zero-frequency susceptibility (and mobility), the
frequency-dependent mobility can be evaluated by an additional (numeric)
integration\cite{Feynman1962}, and the optical response of the polaron
calculated\cite{Devreese1972}.

It will be interesting to see whether a more sophisticated method of calculating
electron-phonon matrix elements from a band-structure\cite{Giustino2017} can
complement the use of the simple Fr\"ohlich dielectric $\alpha$ parameter. 

The Hellwarth et al.\cite{Hellwarth1999} method of deriving a single effective
polar mode could be supplemented by an explicit calculation of contributions
per mode.
This would allow for some of the structure in the low-temperature mobility
(smoothed-out by the choice of a single effective mode) to return. 

\begin{acknowledgments}
We thank Piers Barnes for discussion and a critical reading of a draft of this
manuscript.
We thank Robert Lovrincic for discussion and bringing to our attention his prior
work\cite{Sendner2016}. 
We acknowledge membership of the UK's HPC Materials Chemistry Consortium, which
is funded by EPSRC grant EP/F067496.  
J.M.F. is funded by EPSRC grant EP/K016288/1.

\end{acknowledgments}

\section{Supplementary Information}
Computer codes to replicate all data and figures in this publication are
shared via a repository on GitHub\cite{GitHub},
\url{https://github.com/jarvist/PolaronMobility.jl}.

\bibliography{polarons}

\end{document}